# Thickness dependence of diode efficiency in superconducting Fe(Se,Te)/FeTe thin-film heterostructure devices

Kaito Arizono[1], Kenshin Inamura[1], Kouta Kondou[1,2], Yusuke Kobayashi[1], Tsutomu Nojima[3], Jobu Matsuno[1,2], and Junichi Shiogai[1,2,†]

[1]*Department of Physics, The University of Osaka, Toyonaka, Osaka, 560-0043, Japan.*

[2]*Division of Spintronics Research Network, Institute for Open and Transdisciplinary Research Initiatives, The University of Osaka, Suita, Osaka, 565-0871, Japan.*

[3]*Institute for Materials Research, Tohoku University, Sendai, Miyagi, 980-8577, Japan.*

[†]The author to whom correspondence should be addressed.

junichi.shiogai.sci@osaka-u.ac.jp

**Abstract**

The superconducting diode effect (SDE) is a nonreciprocal transport phenomenon, in which the superconducting critical current density depends on the polarity of the current. It has attracted recent attention because of its potential applications to a rectifier without energy dissipation. While SDE has been observed in a wide range of superconducting materials with broken inversion symmetry as well as thin-film heterostructures, the microscopic origin linking structural inversion asymmetry of electronic band, spin-orbit interaction, and vortex pinning remains to be clarified. In this study, we investigate SDE in Fe(Se,Te)/FeTe heterostructure devices as a function of the superconducting Fe(Se,Te) layer thickness $t_{FST}$ to elucidate the role of structural inversion asymmetry on the vortex-induced SDE. We find that the SDE efficiency monotonically increases with increasing $t_{FST}$, which can be understood by considering that the band bending in the bulk Fe(Se,Te) layer induces the structural inversion asymmetry and thus, the Rashba spin-orbit interaction. In addition, we demonstrate almost 100% rectification for the Fe(Se,Te)/FeTe heterostructure devices in half- and full-wave oscillation configurations. Our findings point out the importance of structural architecture for realization of highly efficient SDE devices based on superconducting thin-film heterostructures.

## I. INTRODUCTION

The superconducting diode effect (SDE), non-reciprocal response of superconducting critical current density ($J_c$) with respect to the current direction, has been attracting interest for its potential applications to superconducting circuits [1-3] and wireless charging to superconducting magnet [4,5]. The rectification of superconducting current density is characterized by the SDE efficiency $\eta$ given as,

$$\eta = (J_c^+ - J_c^-)/(J_c^+ + J_c^-), \qquad \text{Eq. (1)}$$

with $J_c^+$ and $J_c^-$ being the critical current density for positive and negative directions. Up to now, nonreciprocal transport of the supercurrent is recognized as generic phenomena in inversion-asymmetric superconductors such as non-centrosymmetric layered materials [6,7] or gated surfaces [6,8], thin-film heterostructures of iron-based superconductors [9-12] and cuprates [5,13], as well as in single-element superconductors with asymmetrically-patterned nanostructures [14,15]. Depending on the materials and systems, the mechanism of the SDE has been discussed in terms of asymmetric vortex pinning in bulk or surface barriers [12,14-17] and intrinsic origins related to spin-orbit interaction (SOI) [3,6,18,19].

Figures 1(a) and 1(b) illustrate the mechanism of vortex-induced SDE in the presence of SOI [9]. With structural inversion asymmetry (SIA) along out-of-plane direction (// $z$-axis), the helical spin-split Fermi surface is formed by Rashba SOI ($H_{\mathrm{R}} = \alpha\langle\mathbf{E_z}\rangle(\mathbf{k}\times\hat{\mathbf{z}})\cdot\boldsymbol{\sigma}$). Here, $\alpha_{\mathrm{R}} = \alpha\langle\mathbf{E_z}\rangle$ is Rashba parameter due to electric field $\mathbf{E_z}$ along $\hat{\mathbf{z}}$ direction, and $\alpha$, $\mathbf{k}$ and $\boldsymbol{\sigma}$ represent effective mass parameter, wavenumber vector and Pauli spin matrices, respectively [20,21]. Under the in-plane magnetic field **B** (// $y$-axis), the spin-split band is shifted from its center due to the combination of Zeeman effect ($H_{\mathrm{Z}} = g\mu_{\mathrm{B}}\mathbf{B}\cdot\boldsymbol{\sigma}$) and Rashba SOI as shown in Fig. 1(a). Here, $g$ and $\mu_{\mathrm{B}}$ represent $g$-factor and Bohr magneton. In the superconducting state, the singlet Cooper pair acquires a finite center-of-mass momentum $\Delta\mathbf{q}$. At small $B$ = $|\mathbf{B}|$, $\Delta\mathbf{q}$ in the equilibrium state is given by [18],

$$\Delta\mathbf{q} \sim 2\alpha_{\mathrm{R}}\,(\mathbf{B}\times\hat{\mathbf{z}})/v_{\mathrm{F}}^{2} \qquad \text{Eq. (2)}$$

and proportional to $\alpha_{\mathrm{R}} = \alpha\langle\mathbf{E_z}\rangle$ and $B$. Here, $v_{\mathrm{F}}$ represents Fermi velocity. Our previous study in the Fe(Se,Te)/FeTe heterostructure device pointed out that when $\Delta\mathbf{q}$ is superimposed on the rotational current $\mathbf{J}_{\mathrm{r}}$ around vortices, asymmetric pinning potential is formed as shown in Fig. 1(b) [9]. According to the Josephson's relation, the vortex motion with its velocity **v** along the Lorentz force **J** × **B** (// $z$-axis) generates the electric field **E** = **B** × **v** along the applied current **J** (// $x$-axis). When the pinning potential, and thus **v**, is asymmetric, this induced **E** can be also asymmetric with respect to the polarity of **J**, resulting in SDE.

Tellurium-doped iron selenide Fe(Se,Te), one of 11-type iron-based superconductors, is advantageous not only for in-depth understanding of the microscopic origin of SDE but also for high-field applications because of tunable SOI with Te concentration [22] and their relatively large critical current $J_c$ under the magnetic field [23]. We have recently demonstrated a scaling relation of SDE efficiency in Fe(Se,Te)/FeTe heterostructure device as a signature of the vortex-induced SDE [9]. Despite the fact that crystal structure of Fe(Se,Te) and FeTe is PbO structure ($P4/nmm$) and centrosymmetric, the observation of such vortex-induced SDE coupled with SOI implies the significant role of SIA in the Fe(Se,Te)/FeTe thin-film heterostructure [9]. Thus, the link between SIA of the thin-film heterostructure and the value of $\eta$ should be addressed to understand the microscopic origin of the vortex-induced SDE as well as to optimize the stacking structure for highly efficient SDE devices.

In this study, we investigated the Fe(Se,Te)-layer-thickness ($t_{\mathrm{FST}}$) dependence of $J_c^+$ and $J_c^-$, and examined $\eta$ in $t_{\mathrm{FST}}$-nm-thick Fe(Se,Te)/20-nm-thick FeTe heterostructure devices [Fig. 1 (c)]. We found that $\eta$ monotonically increases with increasing $t_{\mathrm{FST}}$, which can be understood based on the Rashba SOI induced by SIA due to electronic band bending in the Fe(Se,Te) layer. In addition, using the heterostructure device with $t_{\mathrm{FST}}$ = 60 nm showing the largest $\eta$, we demonstrate the almost 100% rectification of output voltage up to 2.5 kHz in half- and full-wave oscillation configurations.

## II. EXPERIMENTAL DETAILS

Se-capped $t_{FST}$-nm-thick Fe(Se,Te)/20-nm-thick FeTe heterostructures were grown on $CaF_2$(001) substrates using pulsed-laser deposition (PLD). The $t_{FST}$ varies as 10, 23, and 60 nm by growth duration. The Fe(Se,Te) and FeTe layers were deposited at growth temperature of 300°C, and the Se cap was deposited at room temperature. All layers were deposited in vacuum with a base pressure of $10^{-6}$ Torr. The PLD targets of Fe(Se,Te) with Se/Te ratio of 1 and FeTe are commercially available (K&R Creation Co. Ltd.) and that of Se was fabricated in-house by pressing Se powders (Osaka Asahi Co., Ltd). The *c*-axis-oriented growth was confirmed for the three samples using x-ray diffraction patterns with Cu-$K\alpha_1$ X-ray source ($\lambda$ = 1.5406 Å) (see Supplementary Materials). Thickness of the films was estimated from Laue fringe patterns of thick single-layer reference samples of Fe(Se,Te) and FeTe. After the growth, the Fe(Se,Te)/FeTe heterostructures were patterned into rectangular-shaped channels using water lift-off technique and four-terminal electrodes were deposited by photolithography, Ti/Pt sputtering, and lift-off technique for electrical transport measurements. Details of the sample fabrication are given in Ref. [9]

Optical micrograph of the final device and the measurement setup are provided in Fig. 1(d). The superconducting transition temperature ($T_c$) was estimated from temperature dependence of the sheet resistance measured using current source (Keithley 6221) and lock-in amplifier (SRS830, Stanford Research Systems). Critical current $I_c$ was obtained from current-voltage ($IV$) characteristics measured using a current source (Keysight B2901A) and a nanovoltmeter (Keithley 2182A), and critical current density $J_c$ was calculated by $I_c/(wt_{FST})$ with $w$ being channel width. Data for $t_{FST}$ = 23 nm was obtained from the identical sample presented in our previous work [9]. Temporal variations of input current and output voltages in half- and full-wave oscillation configurations were measured using oscilloscopes. All measurements were performed in a variable temperature insert equipped with a 15 T superconducting magnet (Oxford Instruments, plc).

## III. RESULTS AND DISCUSSION

### A. Dependence of superconducting critical parameters on Fe(Se,Te) layer thickness

Figure 2(a) shows the temperature ($T$) dependence of the sheet resistance ($R_{xx}$) for all samples, which exhibit a clear superconducting transition. Here, transition temperature ($T_c$) is defined as the temperature corresponding to the midpoint of the superconducting transition (dashed lines). For $t_{FST}$ = 60 nm, $T_c$ reaches 14.5 K, which is comparable to the bulk value of $FeSe_{0.5}Te_{0.5}$ [24,25,26]. As $t_{FST}$ decreases to 23 and 10 nm, $T_c$ decreases to 12.8 and 9.0, respectively. One possibility is the presence of epitaxial strain from $CaF_2$ substrate. In Fe(Se,Te) systems, it was revealed that the $c$-axis length and $T_c$ have a linear relation [27]. The thin film of Fe(Se,Te) (in-plane lattice constant: $a$ = 3.818 Å [26]) is grown on $CaF_2$ ($a/\sqrt{2}$ = 3.862 Å) with 45° in-plane rotation [28]. Therefore, the Fe(Se,Te) layer is subjected to the in-plane tensile strain of 2% and the $c$-axis length is shortened. As a consequence, $T_c$ for $t_{FST}$ = 23 nm is reduced from that for the fully-relaxed 60-nm-thick film as shown in Fig. 2(b).

Despite the $c$-axis length for $t_{FST}$ = 10 nm is comparable to that for $t_{FST}$ = 23 nm, $T_c$ further decreases and transition width [the $R(T)$ slope around $T_c$] becomes broadened, which may be ascribed to decrease of superfluid density in two dimensions (2D) by thermal and phase fluctuation. To address dimensionality of superconductivity in the 10-nm-thick Fe(Se,Te) film, we analyze $R(T)$ and $IV$ characteristics at zero in-plane magnetic field ($B$) based on Berezinskii–Kosterlitz–Thouless (BKT) transition [29-31]. Below the BKT transition temperature, vortex-antivortex pairs are bound, resulting in zero resistance. With increasing $T$, the finite resistance emerges by the fluctuation, following Halperin-Nelson formula,

$$R_{\mathrm{xx}} = R_0 \exp\left[-b\left(T/T_{\mathrm{BKT}}^{\mathrm{RT}} - 1\right)^{-1/2}\right]. \qquad \text{Eq. (3)}$$

This leads to

$$\left(\frac{\mathrm{d}\log(R_{\mathrm{xx}}/R_0)}{\mathrm{d}\,T}\right)^{-2/3} \sim \left(T - T_{\mathrm{BKT}}^{\mathrm{RT}}\right). \qquad \text{Eq. (4)}$$

Here, $T_{\mathrm{BKT}}^{\mathrm{RT}}$ is transition temperature. As shown in Fig. 2(c), $R_{\mathrm{xx}}(T)$ curve obeys Eq. (4) and yields $T_{\mathrm{BKT}}^{\mathrm{RT}}$ = 5.60 K. According to the BKT theory, the Lorentz force, generated by a small current, causes out-of-plane vortex-antivortex pairs to unbind, and $IV$ characteristics exhibit a power law $V \propto I^{\alpha}$ showing $\alpha$ = 3 at the BKT transition temperature. Actually, the $IV$ characteristics for $t_{\mathrm{FST}}$ = 10 nm exhibits the BKT-like power-law dependence and the BKT transition temperature $T_{\mathrm{BKT}}^{\mathrm{IV}}$ resides between 5 to 6 K, as shown in Fig. 1(d). The agreement between $T_{\mathrm{BKT}}^{\mathrm{RT}}$ and $T_{\mathrm{BKT}}^{\mathrm{IV}}$ indicates the appearance of 2D superconductivity for this sample.

Figure 2(e) shows $B$ dependence of the superconducting critical current density ($J_{\mathrm{c}}$) at $T$ = 4.2 K. Here, $J_{\mathrm{c}}$ is determined when the $V$ of 100 μV is generated in $IV$ characteristics. For $t_{FST}$ = 60 and 23 nm, $J_c$ at $T$ = 4.2 K and $B$ = 0 T reaches $3 \times 10^5$ $Acm^{-2}$ being comparable to bulk $FeSe_{0.5}Te_{0.5}$ single crystal [9,22,]. Note that, for $t_{FST}$ = 60 nm, $J_c(B)$ is suppressed by only one-third from $J_c(B = 0)$ despite the application of $B$ as large as 15 T. Such a strong $J_c$ against $B$ is characteristic of iron-based superconductors [22]. Figure 2(f) shows $T$ dependence of $J_c$ in $B$ = 0 T, where $J_c(T)$ rapidly decreases with increasing $T$ towards $T_c$ for all the samples.

**B. Structural optimization of superconducting diode efficiency**

Figures 3(a) - 3(c) show typical $IV$ characteristics for $t_{FST}$ = 60, 23, and 10 nm, respectively, at $T \sim 0.5T_c$ under in-plane $B$. For $t_{FST}$ = 60 and 23 nm, the critical current for the positive bias is larger than that for the negative bias when $B > 0$. This relationship is reversed when $B < 0$. Such dependences of critical current on the polarity of bias and $B$ are typical characteristics of SDE. In contrast for $t_{FST}$ = 10 nm, the $IV$ characteristics for positive and negative $B$ almost coincides. Figures 3(d) and 3(e) show $B$ dependence of $J_c$ for a positive and negative bias (denoted as $J_c^+$ and $J_c^-$) and extracted diode efficiency ($\eta$) using Eq. (1). For fair comparison among the samples with different $T_c$, $J_c$ and $\eta$ are compared at nearly the same normalized temperature $T \sim 0.5T_c$. The $\eta(B)$ for $t_{SFT}$ = 23 and 60 nm show typical antisymmetric dependence, exhibiting a linear

dependence in a low $B$ region, reaching maxima, and then monotonically decreasing with further increase of $B$. The maximum value of $|\eta|$, denoted as $\eta_{\mathrm{max}}$, is strongly dependent on $t_{\mathrm{FST}}$: $\eta_{\mathrm{max}}$ reaches 10% at $B = 5$ T for $t_{\mathrm{FST}} = 23$ nm, being consistent with our previous work performed at $T = 4.2$ K [9], while it increases to 17% at $B = 8$ T for $t_{\mathrm{FST}} = 60$ nm.

We discuss the $t_{\mathrm{FST}}$ and $B$ dependences of $\eta$ in the framework of coexistence of the Rashba SOI due to SIA and the Zeeman effect. Figure 3(f) illustrates the schematic profile of conduction band minimum (CBM), $\varphi(z)$, in the $t_{\mathrm{FST}}$-nm-thick Fe(Se,Te)/20-nm-thick FeTe heterostructures under the assumption that $\varphi(z)$ is continuously connected at the Fe(Se,Te)/FeTe interface. Here, $\varphi$ is electron energy. While the Fe(Se,Te) compounds possess a semi-metallic electronic band structure with CBM at M-point and valence band maximum at Γ-point [32,33], we show only CBM at M-point as $\varphi(z)$ for simplicity. Consider CBM is bent upwards by the surface potential at the boundary with the insulating Se cap in such a semimetal with small Fermi pockets [34], non-zero internal electric field $\mathbf{E_z}$ characterized by $\mathrm{d}\varphi/\mathrm{d}z$ is generated in the region $W_{\mathrm{b}}$. When $t_{\mathrm{FST}}$ is thick enough compared to $W_{\mathrm{b}}$, the band bending *i.e.*, $\mathbf{E_z}$ is extended to the bulk Fe(Se,Te) layer and thus, the intense Rashba parameter $\alpha_{\mathrm{R}} = \alpha\langle\mathbf{E_z}\rangle$ emerges. As a result, $\eta_{\mathrm{max}}$ is enhanced as observed for $t_{\mathrm{FST}} = 60$ nm. On the other hand, $t_{\mathrm{FST}}$ is comparable or thinner than $W_{\mathrm{b}}$, the band bending is strongly suppressed by band pinning at the boundaries with Se cap and $CaF_2$ substrate, as evidenced by the suppression of SDE for $t_{FST}$ = 10 nm. Note that while the band bending in the FeTe layer produces opposite sign of $\mathbf{E_z}$ to that in the Fe(Se,Te) layer, the $\mathbf{E_z}$ in the Fe(Se,Te) layer is only effective for SDE in such asymmetric superconducting Fe(Se,Te)/non-superconducting FeTe heterostructures.

Last, we consider the suppression of $\eta(B)$ in the high-$B$ region for $t_{FST}$ = 23 and 60 nm. The monotonic increase of $\eta(B)$ by increasing $B$ at low field region for $t_{FST}$ = 23 and 60 nm as shown in Fig. 3(e), is consistent with $B$-linear dependence of $\Delta\mathbf{q}$ given as Eq. (2). At high field region, in contrast, $|\eta(B)|$ decreases with further increase of $B$ leaving its maximum $\eta_{\max}$. The peak and valley structures of $\eta(B)$ can be ascribed to competition between Zeeman effect and SOI. Under a large magnetic field, spin-momentum-locked state depicted in Fig. 1(a) is violated by Zeeman effect and reorientation of spin textures occurs. As a result, singlet Cooper pair with up- and down-spins cannot be formed, leading to suppression of SDE.

**C. Time-domain measurement of rectification in half- and full-wave oscillation configuration.**

Finally, we demonstrate time-domain measurement of the rectification device operation for $t_{FST}$ = 60 nm under the optimal condition of $T$ = 5.5 K and $B$ = ±8 T. Figure 4(a) shows experimental setup of the half-wave oscillation configuration. A constant AC current with peak-to-zero amplitude of 10 mA and variable frequency ($f$) was applied. Time ($t$) evolution of the current passing through the load resistance $R_L$ = 100 Ω and four-terminal voltage across the device were monitored as output voltage $V_1$ and $V_2$ by oscilloscopes. Figure 4(b) shows measured current $I = V_1/R_L$ and voltage $V_2$ as a function of $tf$ for $f$ = 10 Hz (left panel) and 2.5 kHz (right panel). At frequency of 10 Hz, a finite $V_2$ emerges only when $I > 0$ at $B = -8$ T, while a finite $V_2$ emerges only when $I < 0$ at $B = +8$ T. These responses are quantitatively consistent with $IV$ characteristics shown in Fig. 3(a). Such a clear rectification operation is observed up to 2.5 kHz with slight distortion owing to appearance of inductance component ($\pi/2$ phase shift), signaling the limit of our measurement setup. Applying high-frequency cables and wiring to the devices enables investigation on the detailed dynamics of vortex pinning with $f$ ranging up to a few GHz [35,36]. However, we will leave such experiments as a subject for future work.

Figure 4(c) shows $f$ dependence of the positive ($V^+$, closed circles) and negative ($V^-$, open circles) voltage values, defined as the $V_2$ values measured when $I$ reaches positive peak and negative peak when $B = -8$ T, respectively. Figure 4(d) shows the same traces as those in Fig. 4(c) when $B = +8$ T is applied. The values of $V^+$ for $B = -8$ T and $V^-$ for $B = +8$ T are almost constant with respect to the variation of $f$. In contrast, $V^-$ for $B = -8$ T and $V^{+}$ for $B = +8$ T exhibits zero owing to superconductivity up to $f \sim 2.5$ kHz. In such demonstrations, the figure of merit of rectification is defined by measured voltages as $\eta_V = (|V^+| - |V^-|)/(|V^+| + |V^-|)$. Figure 4(e) shows $f$ dependence of $\eta_V$, exhibiting ±99% at $f = 10$ Hz and slightly decreases towards our measurement limit.

We also demonstrate full-wave oscillation rectification in a bridge having four SDE devices schematically shown in Fig. 5(a). The four SDE devices were fabricated from the same blanket film of the 60-nm-thick Fe(Se,Te)/20-nm-thick FeTe heterostructure as shown in Fig. 5(b) and each device was connected by Constantan wires. Figure 5(c) shows time evolution of input current with a peak-to-zero amplitude of 30 mA and measured output voltage at $f = 10$ Hz under $T = 5.5$ K and $B = -8$ T. Almost perfect rectification was clearly observed. The difference in output voltages for positive and negative current polarities stem from in-plane spatial variation of $I_c$ within the film, which remains to be improved in near future by improving homogeneity of thin-film quality and thickness or introducing microfabrication. The demonstration of full-wave rectification with a relatively large current indicates the potential of the SDE device based on Fe(Se,Te) thin films for superconducting high-power applications such as wireless charging and power monitoring [4,5].

## IV. Conclusion

In conclusion, we investigate the dependence of the superconducting diode effect on Fe(Se,Te) layer thickness ($t_{FST}$) in the superconducting Fe(Se,Te)/non-superconducting FeTe asymmetric heterostructures. With increasing $t_{FST}$, superconducting transition temperature and critical current density increase to those of the bulk single crystals. In addition, the superconducting diode efficiency monotonically increases with increasing $t_{FST}$. This behavior can be ascribed to effective band bending in the superconducting layer and corresponding Rashba spin-orbit interaction due to structural inversion asymmetry of the bulk Fe(Se,Te) layer. We also demonstrate almost 100% rectification effect in time-domain device operation with half- and full-wave oscillation configurations for the 60-nm-thick Fe(Se,Te)/20-nm-thick FeTe heterostructure device. Our findings of the critical role of structural inversion asymmetry in the Fe(Se,Te) heterostructure shed light on the importance of the band engineering of the superconducting materials, especially for low-carrier-density superconductors. In addition, the demonstration of superior rectification paves the way for development of high-power superconducting devices.

**Acknowledgements**

The authors thank Yuji Tsuchiya for stimulating discussion. This work was supported by JST, PRESTO Grant No. JPMJPR21A8 and JSPS KAKENHI Grant No. 23K26379. This work was also supported by The Thermal & Electric Energy Technology (TEET) Foundation.

**Figure captions**

Figure 1 | (a) A schematic of helical spin texture at Fermi surface in the presence of Rashba-type spin-orbit interaction and in-plane magnetic field **B**. The Δ**q** represents finite center-of-mass momentum of the Cooper pair. (b) Schematic vortex motion with a velocity **v** along the *z* direction when the transport current density **J** is applied along *x*, **B** is applied along *y*. $\mathbf{J}_r$ and **E** represent rotational current around the vortex core and electromotive force **E** = **B** × **v**. (c) A schematic of device structure and layer geometry of $t_{FST}$-nm-thick Fe(Se,Te)/20-nm-thick FeTe heterostructure. (d) Optical micrograph of the typical device and measurement setup.

Figure 2 | (a) Temperature dependence of sheet resistance for $t_{FST}$-nm-thick Fe(Se,Te)/20-nm-thick FeTe heterostructure devices with $t_{FST}$ = 10 (blue), 23 (green), and 60 nm (orange). Superconducting transition temperature ($T_c$) is defined as midpoint temperature. (b) The *c*-axis length dependence of $T_c$ for this work (closed circles) and previous studies (white squares) [27]. (c) The plot of $(\mathrm{d}\log(R_{xx}/R_0)/\mathrm{d}\,T)^{-2/3}$ vs $T$ for $t_{FST}$ = 10 nm (blue circles) and linear fitting (black) at zero magnetic field ($B$). (d) The log-log plots of the current-voltage characteristics with various measurement temperatures $T$ for $t_{FST}$ = 10 nm at $B$ = 0 T. The black lines are linear fitting curves with slope $\alpha$ and the dotted red

line corresponds to $\alpha = 3$. The inset shows the $T$ dependence of $\alpha$. (e)(f) Critical current density ($J_c$) as functions of (e) $B$ at $T = 4.2$ K and (f) $T$ at $B = 0$ T.

Figure 3 | (a)-(c) IV characteristics for $t_{FST}$-nm-thick Fe(Se,Te)/20-nm-thick FeTe heterostructure devices with (a) $t_{FST} = 60$ nm at magnetic field ($B$) of ±8 T, (b) 23 nm at $B = \pm 6$ T, and (c) 10 nm at $B = \pm 15$ T. (d) Extracted critical current density for positive and negative biases ($J_c^+$ and $J_c^-$) as a function of $B$. (e) Calculated efficiency of superconducting diode effect, $\eta = (J_c^+ - J_c^-)/(J_c^+ + J_c^-)$ as a function of $B$. Measurement temperature $T \sim 0.5T_c$ is used for each device for the sake of fair composition. (f) A schematic profile of conduction band minimum along out-of-plane direction with different $t_{FST} = 60$ (orange), 23 (green), and 10 nm (blue).

Figure 4 | (a) Setup for time domain measurement. (b) Time ($t$) evolution of measured current (black solid line) and voltages when magnetic field $B = -8$ T (blue) and +8 T (red) is applied. The frequency ($f$) of the applied sinusoidal current is 10 Hz (left panel) and 2.5 kHz (right panel). Horizontal black dashed lines indicate $I = 0$ mA or $V = 0$ mV. (c) Frequency dependence of measured voltages when the sinusoidal current reaches a positive peak ($V^+$, closed circles) and a negative peak ($V^-$, open circles) under $B = -8$ T.

(d) The same $V^+$ and $V^-$ traces as those in (c) under $B$ = +8 T. (d) Voltage rectification efficiency defined as $\eta_V = (|V^+| - |V^-|)/(|V^+| + |V^-|)$ as a function of frequency when $B$ = −8 T (blue) and +8 T (red).

Figure 5 | (a) A schematic bridge circuits for full-wave rectification measurement. (b) The optical microscopic image of Fe(Se,Te)/FeTe devices patterned on a substrate. (c) Time evolution of measured input current and output voltage when magnetic field $B$ = −8 T (blue) is applied. Horizontal black dashed lines indicate $I$ = 0 mA or $V$ = 0 mV.

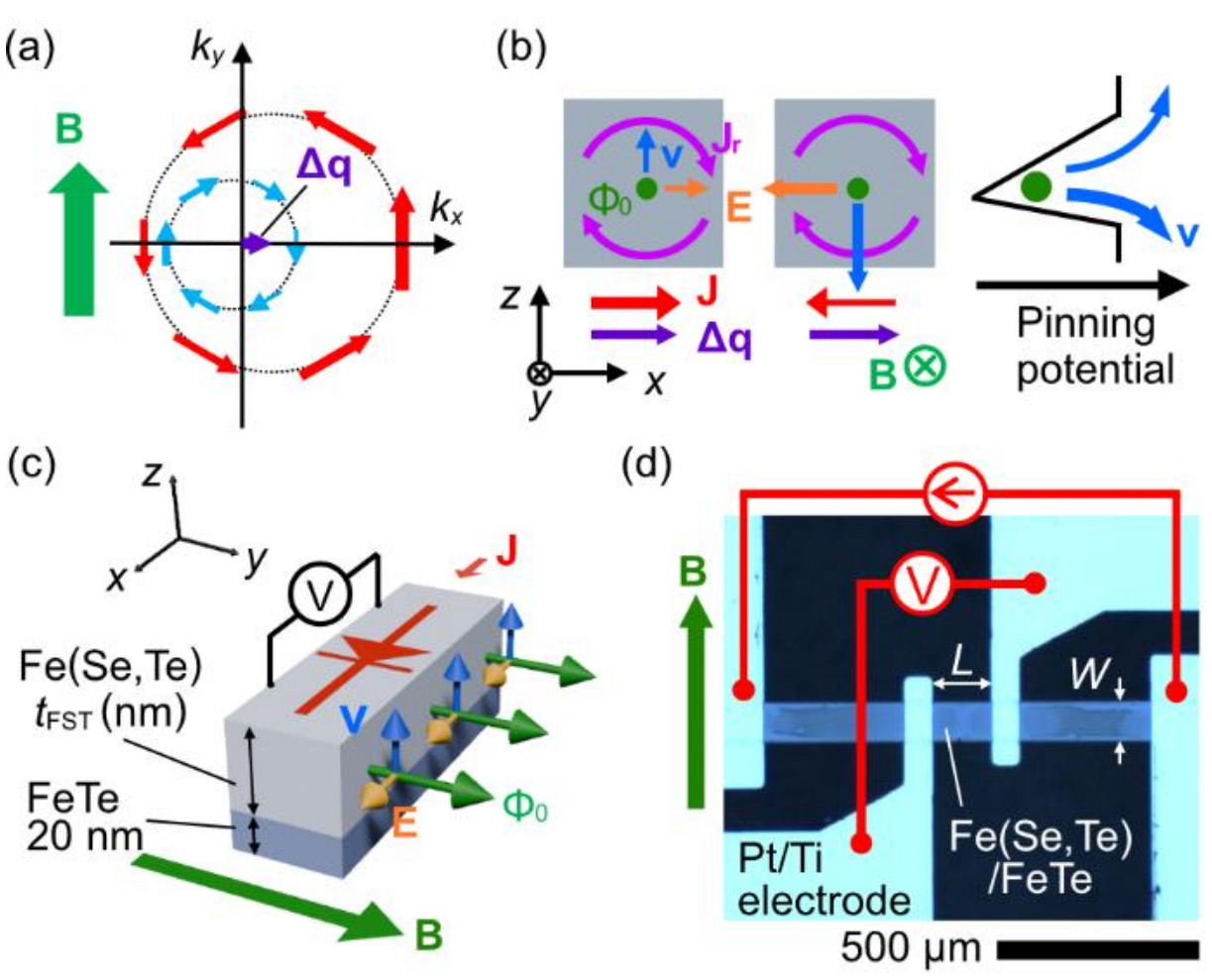

Figure 1 (single column) Arizono et al.

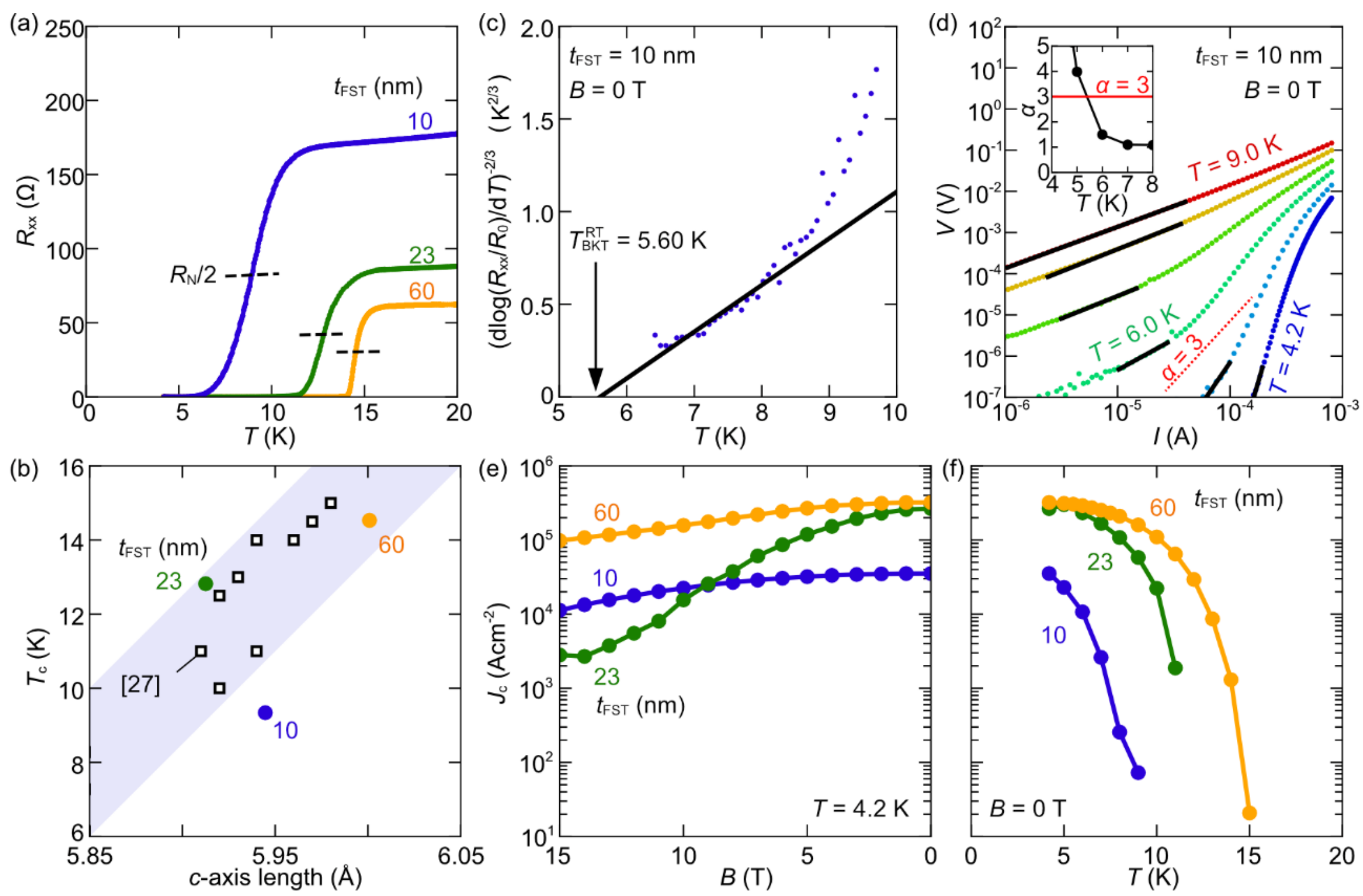

Figure 2 (double column) Arizono et al.

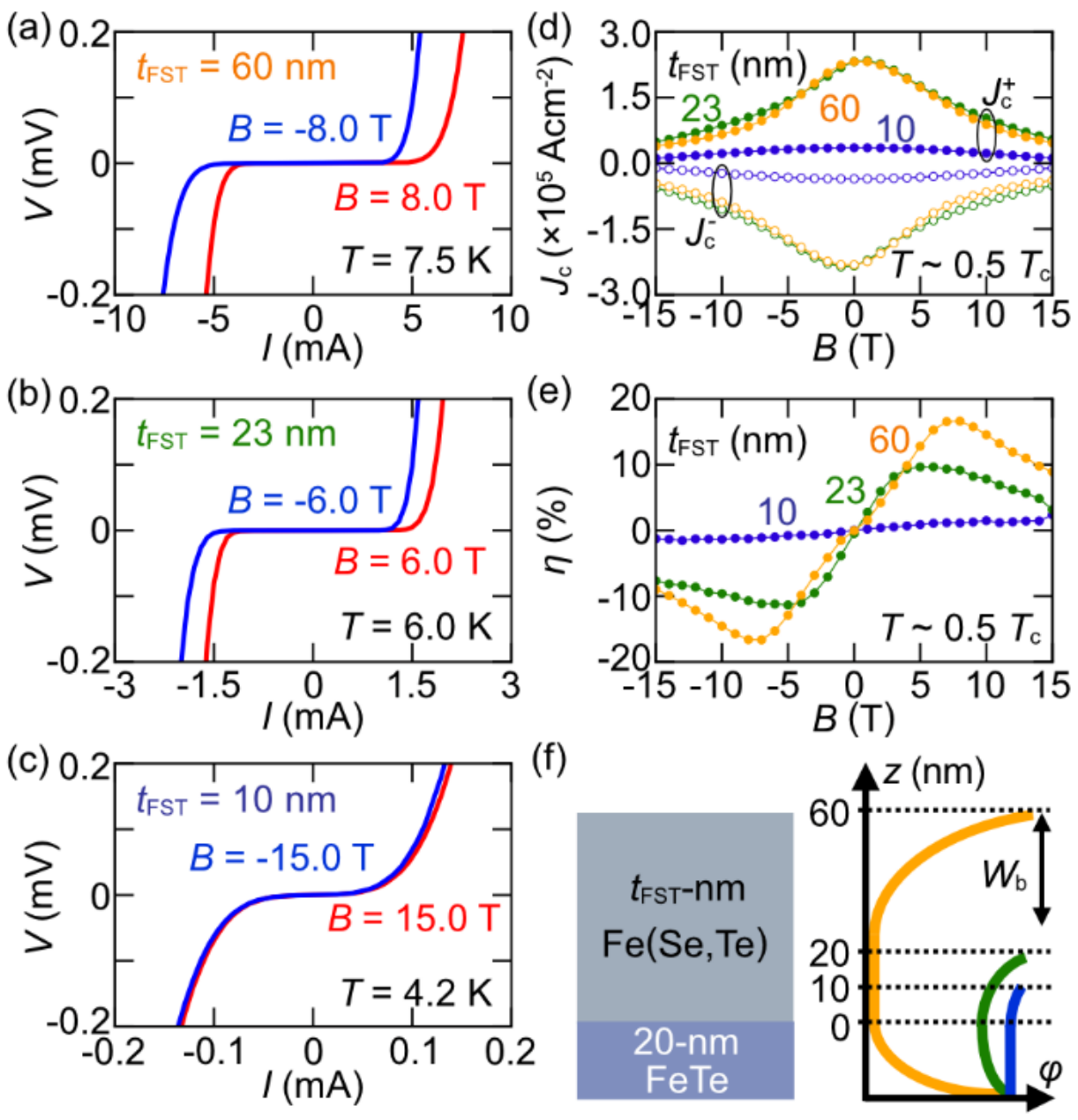

Figure 3 (single column) Arizono et al.

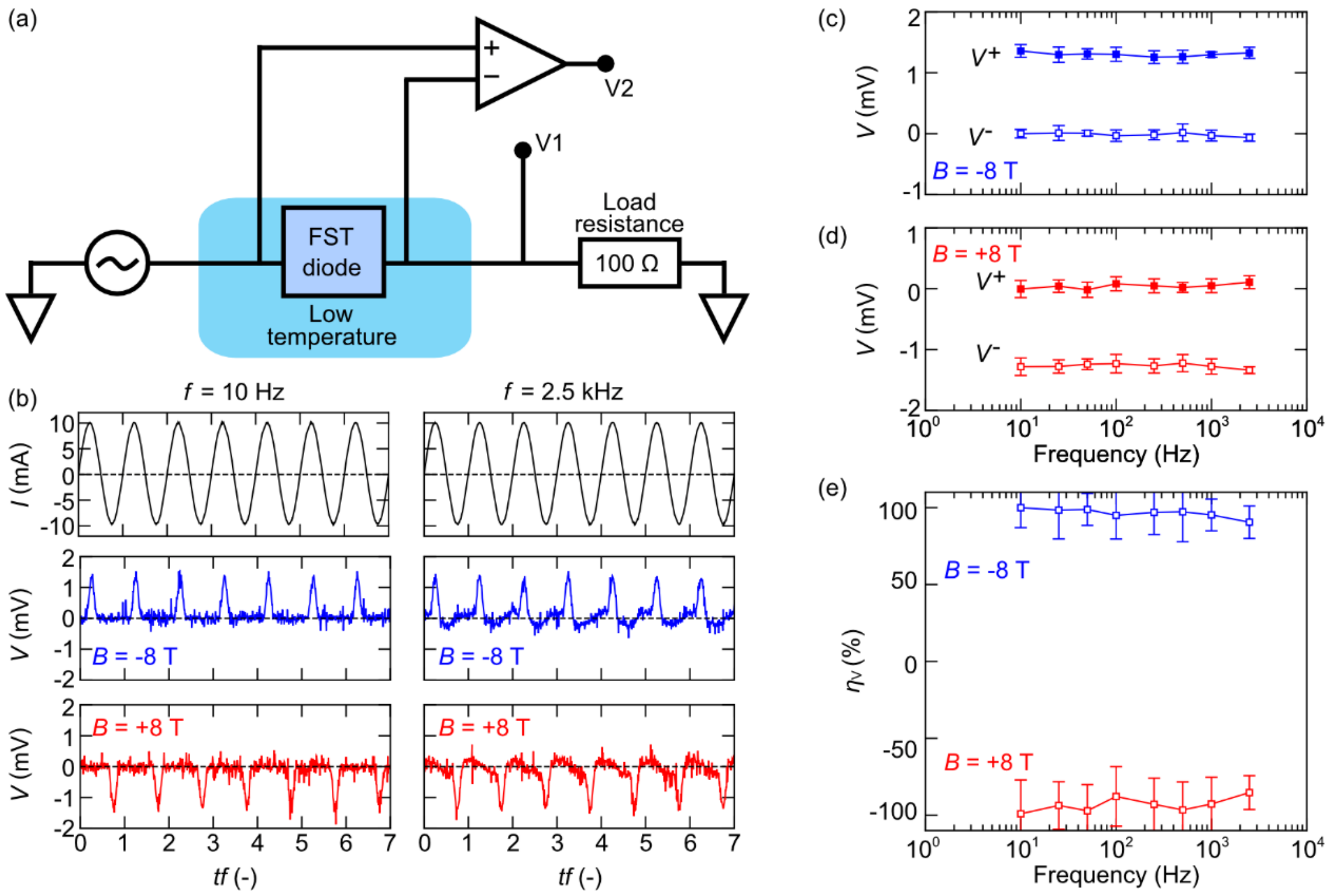

Figure 4 (double column) Arizono et al.

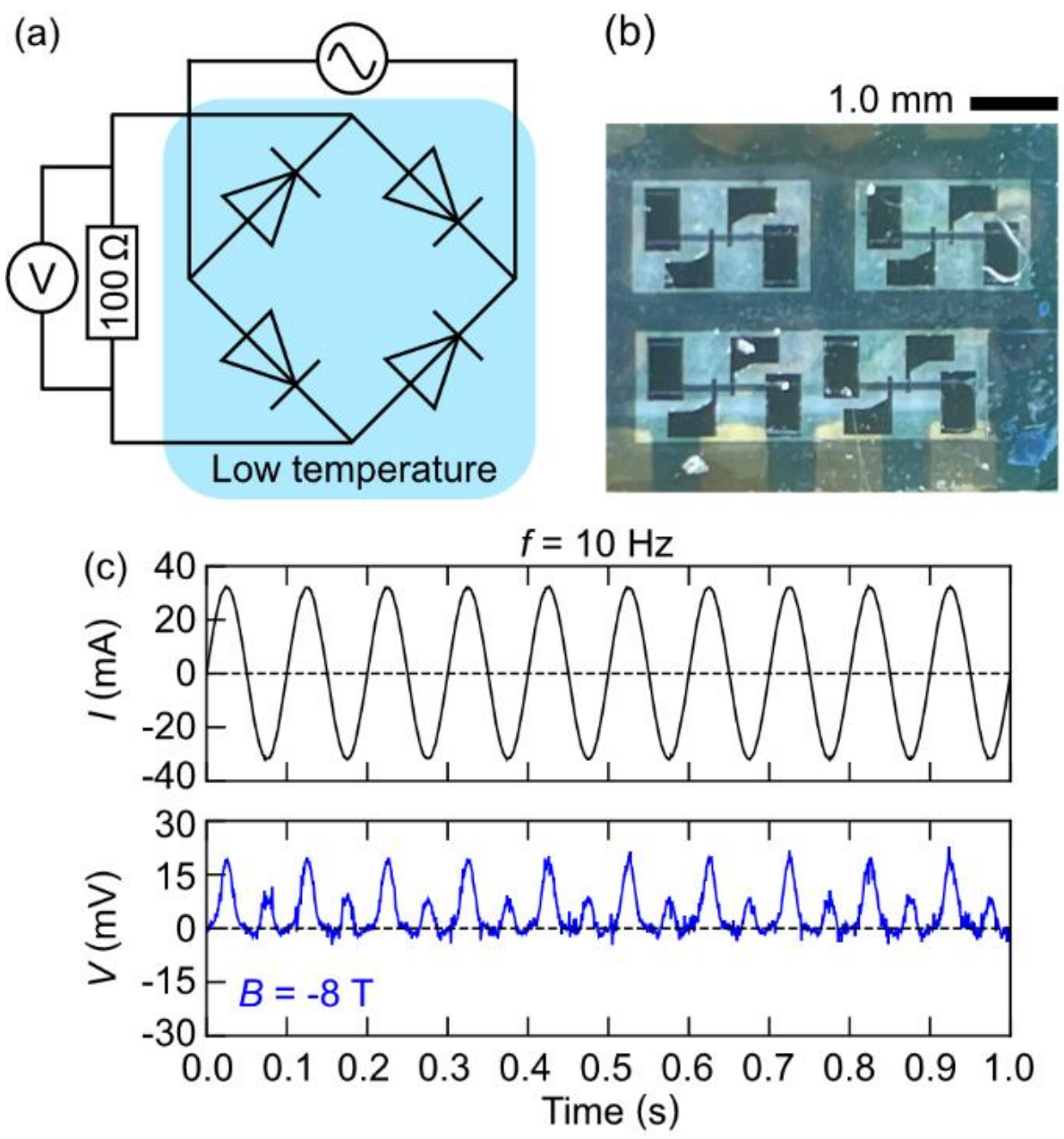

Figure 5 (single column) Arizono et al.

Supplementary Materials for

# Thickness dependence of diode efficiency in superconducting Fe(Se,Te)/FeTe thin-film heterostructure devices

Kaito Arizono[1], Kenshin Inamura[1], Kouta Kondou[1,2], Yusuke Kobayashi[1],

Tsutomu Nojima[3], Jobu Matsuno[1,2], and Junichi Shiogai[1,2]

[1]*Department of Physics, The University of Osaka, Toyonaka, Osaka, 560-0043, Japan.*

[2]*Division of Spintronics Research Network, Institute for Open and Transdisciplinary Research Initiatives, The University of Osaka, Suita, Osaka, 565-0871, Japan.*

[3]*Institute for Materials Research, Tohoku University, Sendai, Miyagi, 980-8577, Japan.*

## 1. Structural characterization using x-ray diffraction pattern.

Structural characterization for Se-capped $t_{FST}$-nm-thick Fe(Se,Te)/20-nm-thick FeTe heterostructures ($t_{FST}$ = 60, 23 and 10 nm) on $CaF_2$(001) substrate was performed by out-of-plane x-ray diffraction (XRD) patterns. Supplementary Figure 1 shows XRD patterns for Fe(Se,Te)/FeTe heterostructures with (a)(b) $t_{FST}$ = 60 nm, (c)(d) 23 nm, and (e)(f) 10 nm. All samples exhibit *c*-axis oriented growth and *c*-axis length is shortened with decreasing $t_{FST}$ by in-plane tensile strain as discussed in the main text.

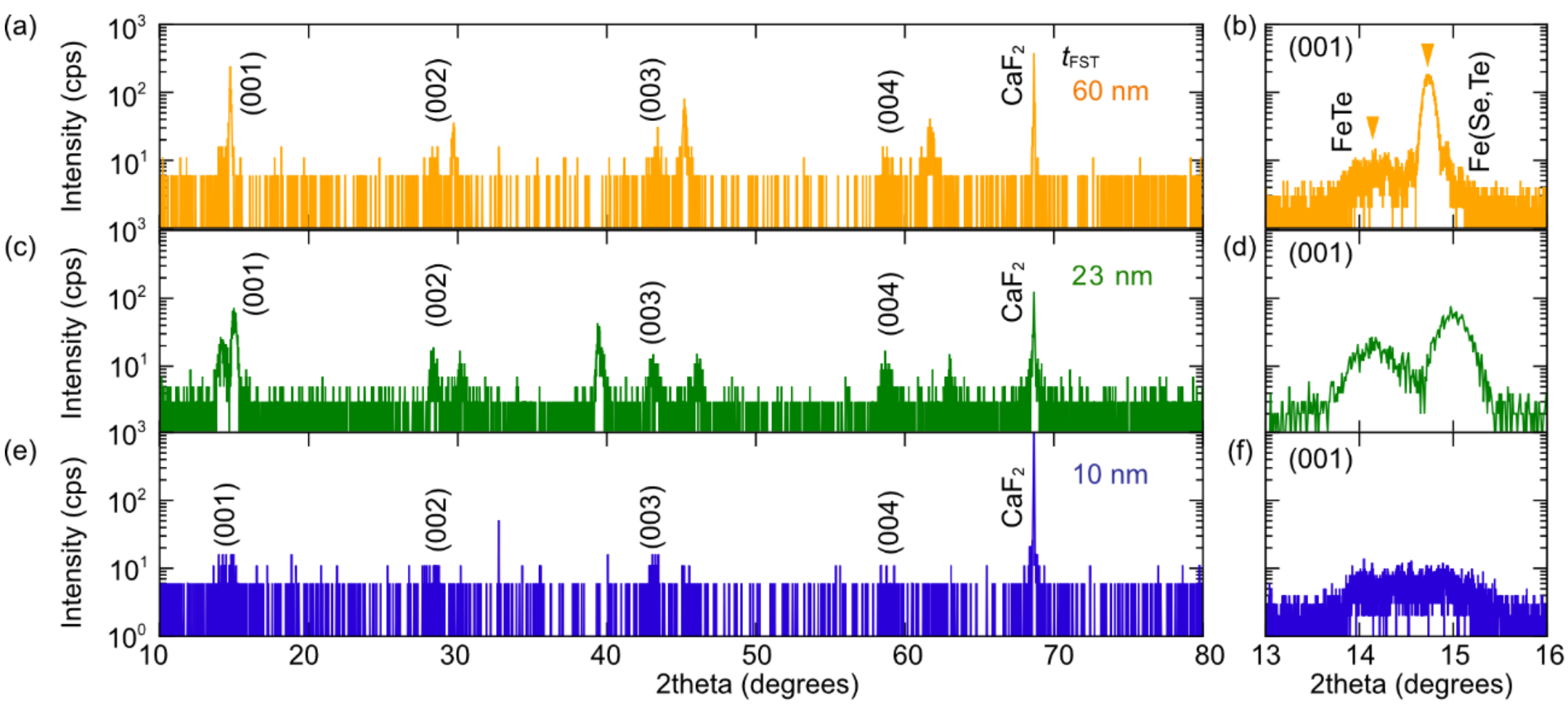

**Supplementary Figure 1 | Structural Characterization.** X-ray diffraction patterns of Fe(Se,Te)/FeTe heterostructures for (a)(b) $t_{FST}$ = 60 nm, (c)(d) 23 nm, and (e)(f) 10 nm. (a), (c) and (e) are wide range of 2theta. (b), (d) and (f) are around (001) diffraction peak of Fe(Se,Te) and FeTe.